\documentclass{aa}
\input epsf
\def\gsim{\ \raise -2.truept\hbox{\rlap{\hbox{$\sim$}}\raise5.truept\hbox{$>$}}}

\begin{document}
%\thesaurus{03(11.01.2; 11.02.1; 11.02.2 3C~66A,ON~325; 11.17.3)}
\title{Log-parabolic spectra and particle acceleration in blazars - II: \\
The BeppoSAX wide band X-ray spectra of Mkn~501}
\author{E.~Massaro\inst{1,2}
\and M.~Perri\inst{3}
\and P.~Giommi\inst{3}
\and R.~Nesci\inst{1}
\and F.~Verrecchia\inst{3}
\institute{
Dipartimento di Fisica, Universit\`a La Sapienza, Piazzale A. Moro 2,
I-00185 Roma, Italy
\and IASF - Sezione di Roma, INAF-CNR, via del Fosso del Cavaliere,
I-00113 Roma, Italy
\and ASI Science Data Center, ESRIN, I-00044 Frascati, Italy
}}
\offprints{enrico.massaro@uniroma1.it}
\date{Received ....; accepted ....}

\markboth{E. Massaro et al.: The log-parabolic X-ray spectra of Mkn~501}
{E. Massaro et al.: The log-parabolic X-ray spectra of Mkn~501}

\abstract{
We present the results of a spectral and temporal study of the complete set of 
{\it Beppo}SAX NFI (11) and WFC (71) observations of the BL Lac object Mkn~501.
The WFC 2--28 keV data, reported here for the first time, were collected over 
a period of about five years, from September 1996 to October 2001.
These observations, although not evenly distributed, show that Mkn~501, 
after going through a very active phase from spring 1997 to early 1999, 
remained in a low brightness state until late 2001.
The data from the LECS, MECS and PDS instruments, covering the wide energy 
interval 0.1--150 keV, have been used to study in detail the spectral 
variability of the source. 
We show that the X-ray energy distribution of Mkn~501 is well described by a
log-parabolic law in all luminosity states.
This model allowed us to obtain good estimates of the SED synchrotron peak 
energy and of its associated power.
The strong spectral variability observed, consisting of strictly correlated 
changes between the synchrotron peak energy and bolometric flux 
($E_{peak} \propto F_{b}^{\sim 1.4}$), suggests that the main physical changes 
are not only due to variations of the maximum Lorentz factor of the emitting 
particles but that other quantities must be varying as well. 
During the 1997 flare the high energy part of the spectrum of Mkn~501
shows evidence of an excess above the best fit log-parabolic law suggesting the
existence of a second emission component that may be responsible for most of the 
observed variability.
}
\authorrunning{E. Massaro et al.}
\titlerunning{The log-parabolic X-ray spectra of Mkn~501}

\maketitle

\section{Introduction}

The study of the wide band Spectral Energy Distributions (SEDs) of BL Lac objects
(and Blazars in general) has shown that these sources are characterized by a double
luminosity peak structure. The peak at lower energies is generally explained by
synchrotron radiation from relativistic electrons in a jet closely aligned to the
line of sight, while the high frequency bump is produced by inverse Compton scattering.
The peak frequency of the first bump ranges from the Infrared-Optical region for the
so called Low-energy peaked BL Lac (LBL) objects to the UV-X ray range for
the High-energy peaked BL Lac (HBL) (Padovani and Giommi 1995).
The shape of these bumps is characterized in the $Log(\nu F_{\nu})$ vs. $Log~\nu$ plots
by a rather smooth curvature extending through several frequency decades.
Analytical models have been proposed to represent these SEDs: a very simple
and successful model is a log-parabola with only three spectral parameters.
In a previous paper (Massaro et al. 2004, hereafter Paper I) we used the log-parabolic
model to fit the {\it Beppo}SAX wide band X-ray spectra of the BL Lac object Mkn~421.

In this paper we apply the same spectral analysis to the entire data set of 
{\it Beppo}SAX X-ray observations of Mkn~501 which, like Mkn~421, is an HBL source 
and has been detected in the TeV range (Quinn et al. 1996). 
The X-ray luminosity peak has been observed to be very variable and the energy
of the maximum can be as high as about 100 keV (Pian et al.~1998). 
Mkn~501 ($z$=0.0337) has been one of the main targets of multifrequency observational 
campaigns from ground and space observatories like ASCA, RXTE and {\it Beppo}SAX. 
Among all these observations, those of spring 1997 are particularly interesting 
because they were taken during a very bright flare when Mkn~501 reached an integral flux 
in the TeV band of about 10 times that of the Crab Nebula (see, for instance, Aharonian 
et al. 1999, 2001). 

\begin{table*}
\caption{ {\it Beppo}SAX NFI observation log of Mkn~501.}
\label{tab1}
\begin{tabular}{lcccccc}
\hline
 Date & Start UT & End UT & LECS Exposure & MECS Exposure & PDS Exposure & PDS Count Rate$^1$ \\
      &          &        &    (s)        &  (s)          &    (s)       &   (cts/s) \\
\hline
1997/04/07     & 05:11   & 16:02 & ~12,955  & ~20,665 & ~8,936 & 1.94$\pm$0.07 \\
1997/04/11     & 05:22   & 16:26 & ~13,059  & ~20,430 & ~8,720 & 2.35$\pm$0.07 \\
1997/04/16     & 03:20   & 14:36 & ~~9,812  & ~17,124 & ~7,347 & 7.66$\pm$0.08 \\
               &         &       &          &         &         & \\
1998/04/28-29  & 08:52   & 02:19 & ~13,790  & ~21,911 & ~9,873 & 1.50$\pm$0.06 \\
1998/04/29-30  & 20:33   & 13:40 & ~14,917  & ~21,426 & ~9,662 & 1.80$\pm$0.06 \\
1998/05/01-02  & 19:02   & 10:38 & ~13,402  & ~19,045 & ~8,447 & 0.90$\pm$0.07 \\
1998/06/20-21  & 10:33   & 03:57 & ~17,003  & ~25,871 & ~11,566 & 0.43$\pm$0.04 \\
1998/06/29-30  & 19:41   & 12:48 & ~12,939  & ~18,943 & ~7,471 & 0.52$\pm$0.07 \\
1998/07/16-17  & 19:38   & 11:06 & ~11,443  & ~15,942 & ~6,918 & 0.63$\pm$0.07 \\
1998/07/25-26  & 15:19   & 16:52 & ~25,531  & ~30,952 & ~14,505 & 0.40$\pm$0.05 \\
               &         &       &          &         &         & \\
1999/06/10-16  & 23:01   & 02:11 & 105,849  & 175,396 & 78,415 & 0.10$\pm$0.02 \\
\hline
\multicolumn{7}{c} { }
\end{tabular}

$^1$ 15--90 keV energy band.
\end{table*}

Our spectral analysis is based on all the {\it Beppo}SAX observations performed with three 
of the Narrow Field Instruments (NFIs) on-board this satellite: LECS (0.1--10 keV) 
(Parmar et al. 1997), MECS (1.4--10 keV) (Boella et al. 1997) and PDS (13--300 keV) 
(Frontera et al. 1997).
Some of these observations have already been analysed by other authors, but
with different approaches and applying different spectral laws.
Pian et al. (1998) presented the data of the 1997 {\it Beppo}SAX campaign on Mkn~501 
and modelled the spectra with a broken power law. Further analyses of
the 1997-1999 observations were presented by Tavecchio et al. (2001), but four 
observations, performed in June and July 1998, were not included in their work.

In this paper we considered all the 11 NFI {\it Beppo}SAX pointed observations of 
Mkn~501 to verify if a log-parabolic law gives a good description 
of the synchrotron peak in the SED of this source.
Furthermore, we analysed all the serendipitous detections of Mkn~501 with 
the two Wide Field Cameras (WFCs) on board {\it Beppo}SAX (Jager et al. 1997).
These two detectors were coded mask cameras operating in the energy range 
2--28 keV with a very large field of view ($40^\circ \times 40^\circ$) but with 
a reduced sensitivity compared to the NFIs. 
These data are useful to obtain information of the long time evolution of the
X-ray luminosity of Mkn~501 over the {\it Beppo}SAX lifetime.
 
In the first part of the paper we present the best fit results of the X-ray observations
and discuss the properties of the Spectral Energy Distribution of Mkn~501, also
taking into account some simultaneous optical data. In the second part we compare 
our results with those of other authors and also with our findings on Mkn~421 (Paper I) 
and discuss some implication of our results on the particle acceleration in the inner jet 
of this source. 

\begin{figure*}
%      \vspace{1.0cm}
       \hspace{1.5cm}
\epsfysize=9cm
\epsfbox{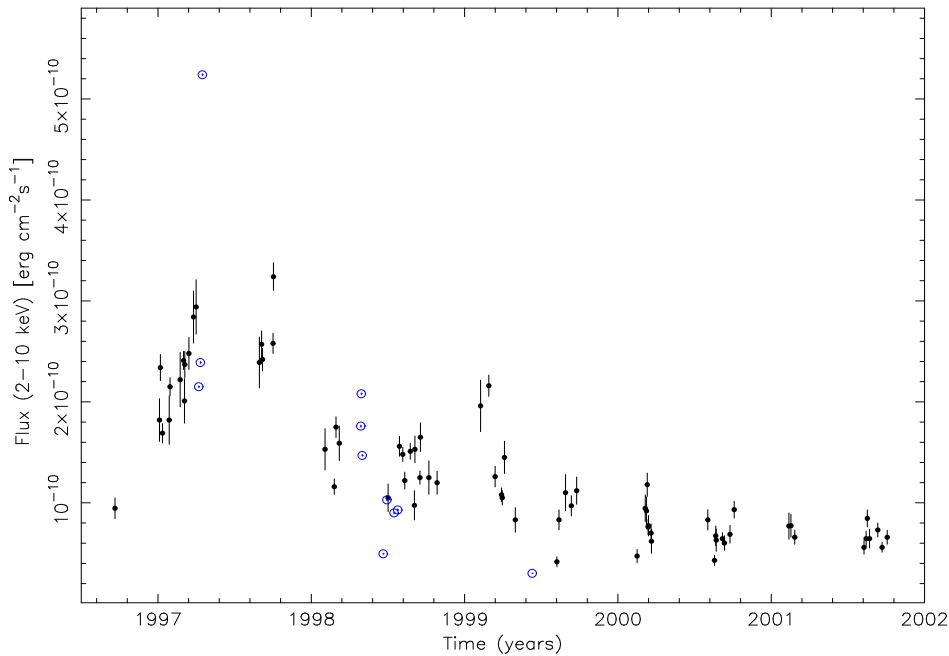}
\caption{{\it Beppo}SAX 2--10 keV light curve of Mkn~501 observed with WFC (filled circles) and 
MECS (open circles). Statistical errors of the latter data are smaller than the 
symbol size.}
%\label{lc210kev}
\end{figure*}

\section{X-ray observations and Data Reduction}

\subsection{NFI Instruments}

{\it Beppo}SAX performed a total of 11 pointed observations of Mkn~501: 3 observations 
in April 1997, 7 in the period from April to July 1998 and the last one in June 1999.
The log of all these observations and the net exposure times for the three
NFIs considered in our analysis are given in Table 1. 
Observations were generally concentrated in time windows of several days and the
typical MECS exposure time was around 20 ks.
A longer observation, with interruptions of only few hours, occurred from 
1998, April 28 to May 2. In 1999 Mkn~501 was pointed again for a much longer time 
and the net exposure was about 175 ks.
We recall that in the observations of April 1997 the MECS operated with all the 
three detectors, while in the subsequent pointings only two detectors were active.

During the observations the count rate of Mkn~501 was generally stable:
the variations, when detectable, did not exceed the 20--30\%. For this
reason and because of the limited statistics we avoided to segment the data in time 
or intensity and analysed the entire data set to achieve a better definition of the
mean spectral shape.

Standard procedures and selection criteria were applied to the data to 
avoid the South Atlantic Anomaly, solar, bright Earth and particle 
contamination using the SAXDAS (v.~2.0.0) package.
Data analysis was performed using the software available in the XANADU 
Package (XIMAGE, XRONOS, XSPEC). The images in the LECS and MECS instruments showed
a bright pointlike source: events for spectral analysis were 
selected in circular regions, centred at the source position, with radii 
of 4$'$ and 8$'$ depending upon the count rate, as indicated by
Fiore et al. (1999).
Background spectra were taken from blank field archive. 

MECS data were always taken in the whole instrumental energy range,
while narrower ranges were used for the LECS and PDS. For the LECS
we limited the upper bound to 2 keV because above this energy the MECS 
provided better statistics and to avoid some problems affecting
the LECS response matrix, while the lower bound was selected
taking into account the signal to noise ratio and the presence of systematic
effects, but it was never higher than 0.2 keV. We choose the upper
bound of the PDS range depending on the source flux and the confusion
limit due to the instrumental field of view. 
As a consequence the useful energy ranges were wider when the source
was brighter. In particular for the long observation of 10--16 June 1999 
PDS data were not used due to the weakness of Mkn~501. 
The PDS count rates in the 15--90 keV of each observation are given in
Table 1, whereas the energy ranges used for the LECS and PDS data are in Table 3.

In the X-ray spectral fitting we considered the low energy absorption 
due to the interstellar gas. The Galactic column density was taken equal to 
$N_H$ = 1.71$\times$10$^{20}$ cm$^{-2}$, derived from the survey by 
Dickey \& Lockman (1990) and in very good agreement with the estimate of
Stark et al. (1992) adopted by other authors (Pian et al. 1998, Sambruna et 
al. 2000, Tavecchio et al. 2001). The higher value of 2.87$\times$10$^{20}$ 
cm$^{-2}$ was derived by Lamer et al. (1996) from the spectral analysis of ROSAT 
data; this estimate, however, could be affected by the intrinsic spectral 
curvature of Mkn~501, not considered by these authors.

\subsection{Wide Field Cameras}

Mkn 501 was detected by the WFCs on 71 occasions from September 1996 to 
October 2001. The log of all these observations is given in Table 2 together with 
the 2--10 keV flux calculated assuming for the source spectrum a power law model with 
photon index $\Gamma = 1.9$. A typical exposure time for the WFC is 30 ks with a few notable 
exceptions like the observation of 17 August 2001 which lasted over five days.

As for NFIs, standard procedures were applied to WFC data in order to select
good time intervals (GTI) for the observations filtering out all unwanted 
contaminations.
Due to the coded mask nature of the instruments, the reconstruction 
of sky images is achieved through an algorithm that cross-correlates the detector 
image with the coded mask structure (Fenimore \& Cannon 1978) in an
iterative process (Hammersley et al. 1992, in 't Zand 1992).

The subsequent data analysis was performed, as for NFIs instruments, using the 
XANADU package. The WFC sensitivity depends mainly on the off-axis angle and on 
the background, so different acceptance criteria on the signal to noise ratio are 
applied to source identifications depending on the pointing direction and on the 
off-axis angle.
This data reduction technique and the data analysis procedure are implemented in the standard 
processing used to build the  WFC science archive and the source 
catalog (Piranomonte et al. 2002, Verrecchia et al. in preparation).\\

The combined MECS and WFC light curve of Mkn~501 in the 2--10 keV band is shown in
Fig.~1: the source went through a very active phase from Spring 1997 to early
1999 and afterwards it remained in a low brightness state. Note also that the very high
flux measured on 16 April 1997 has never been detected in any other observations.
It should be likely considered an exceptional state.

\section{The Spectral analysis}

Wide band X-ray spectral distributions of Mkn~501 are remarkably curved, as
already pointed out by Tavecchio et al. (2001):
the best fits with simple power laws give largely unacceptable $\chi^2$ and 
different analytical models must be considered to describe these 
spectra. We fitted to the data a power law with an exponential cut-off and 
generally obtained high reduced $\chi^2$, in particular, for the 1997
observations when Mkn~501 was very bright at high energies, the reduced $\chi^2$
was never lower than 2. 
Tavecchio et al. (2001) modelled the spectra using a continuous combination 
of two power laws, the same model adopted by Fossati et al. (2000) in the analysis of
Mkn~421.
No restriction is assumed for the values of the two spectral indices and so it
requires four free parameters to be determined. 
Although this model provides good fits, the use of two (or more) spectral indices at 
different energies does not allow a direct and simple measure of the SED curvature.

In our analysis we have instead adopted a log-parabolic model for the spectral distribution:
\begin{equation}
 F(E) = K~ (E/E_1)^{-(a+b~Log(E/E_1))} ~~~{\rm ph/(cm^2~s~keV)}.  
\end{equation}
The main properties of this spectral model are summarized in the following (see Paper I for
details).
We took the reference energy $E_1$ fixed to 1 keV and therefore the 
spectrum is completely determined by the three parameters $K$, $a$ and $b$. 
It is possible to define an energy dependent photon index $\Gamma(E)$, given
by the log-derivative of Eq.(1):
\begin{equation}
 \Gamma(E) = a~ +~ 2~ b~ Log(E/E_1) ~~~~~~~~.  
\end{equation}
The parameter $a$ is then the photon index at the energy $E_1$ and $b$ measures 
the curvature of the parabola: it is easy to demonstrate that the curvature radius
at the parabola vertex is equal to $1/|2b|$. A good estimate of $b$ can be obtained 
only using a sufficiently wide energy range, particularly when its value is small.
We define the peak frequency $\nu_p=E_p/h$, corresponding to the maximum 
in the $Log~\nu F(\nu)$ vs. $Log~\nu$ plot, easily computed from the spectral 
parameters $a$ and $b$, as:
\begin{equation}
 E_p = E_1~ 10^{(2-a)/2b} ~~~~~~~~  
\end{equation}
and 
\begin{equation}
\nu_p F(\nu_p) = (1.60\times10^{-9})~K E_1^2 ~10^{(2-a)^2/4b} ~~{\rm erg/(cm^2~s)}~  
\end{equation}
where the constant is simply the energy conversion factor from keV to erg.

The log-parabolic distribution can be analytically integrated over the entire 
frequency range to estimate the bolometric flux: 
\begin{equation}
 F_{bol} = \sqrt{\pi~ ln10}~ \frac{\nu_p F(\nu_p)}{\sqrt{b}}~ = 
2.70~ \frac{\nu_p F(\nu_p)}{\sqrt{b}} ~~~~. 
\end{equation}
 
The log-parabolic law generally provides a good description of the SED in a wide energy 
interval around the peak. However, at large distances from $E_p$ one can reasonably expect 
deviations from a perfect parabola. In fact, for low particle energies the acceleration 
probability may be energy independent originating a single power law spectrum. On the high 
energy end, the particle spectrum can be limited at the maximum achievable energy producing 
a rather sharp cutoff in the radiation spectrum. Furthermore, the log-parabola is symmetric 
with respect to the peak and therefore cannot describe well highly asymmetric distributions. It is 
easy to modify Eq.~(1) to include one more parameter that takes into account of possible 
asymmetries. 
However, the spectral analysis of the BeppoSAX observations of Mkn~501 did not require the 
addition of a new parameter. 
We then used a log-parabolic law which allowed us to evaluate the mean spectral curvature 
by means only of the parameter $b$.

\begin{table}
\setcounter{table}{2}
\caption{ LECS and MECS energy ranges, reduced $\chi^2$ values and inter-calibration 
parameters for the log-parabolic best fit spectra of Mkn~501.}
\label{tab3}
\begin{tabular}{lccccl}
\hline
Date & LECS & PDS &$\chi^2_r$/dof & $f_{LM}$ & $f_{MP}$ \\
     & keV  & keV &               &          &          \\
\hline
1997/04/07 & 0.2-2 & 15-60 & 1.22/112 & 0.84 & 0.90 f  \\
1997/04/11 & 0.12-2 & 15-130 & 1.05/139 & 0.85 & 0.83 f \\ 
1997/04/16 & 0.12-2 & 15-150 & 1.01/140 & 0.82 & 0.88 f \\
           &        &         &           &      &  \\
1998/04/28-29 & 0.2-2 & 15-90 & 1.10/113 & 0.81 & 0.87 \\
1998/04/29-30 & 0.2-2 & 15-70 & 1.02/112 & 0.83 & 0.88 \\
1998/05/01-02 & 0.2-2 & 15-60 & 1.09/112 & 0.81 & 0.90 f \\
1998/06/20-21 & 0.13-2 & 15-30 &1.00/113 & 0.81 & 0.90 f \\
1998/06/29-30 & 0.12-2 & 15-20 & 1.07/112 & 0.75 & 0.78 f \\
1998/07/16-17 & 0.13-2 & 15-30 & 1.06/113 & 0.75 & 0.90 f \\
1998/07/25-26 & 0.13-2 & 15-45 & 1.09/115 & 0.78 & 0.91 \\
           &        &         &           &     &  \\
1999/06/10-16 & 0.15-2 &    & 1.18/108 & 0.75 &  \\
\hline
\multicolumn{6}{c} { }
\end{tabular}

{\it Note}: f indicates a frozen value.
\end{table}

As stressed in Paper I, it is important, when working with spectra obtained with
two or three NFIs, to use the appropriate inter-calibration factors between MECS 
and LECS ($f_{LM}$) and PDS ($f_{MP}$). 
The accurate ground and the in-flight calibrations were used to establish the
admissible ranges for these two factors which are: 0.7$\leq f_{LM}\leq$1.0 and 
0.77$\leq f_{MP}\leq$ 0.93, the latter reduced to 0.86$\pm$0.03 for sources with a PDS count 
rate higher than about 2 cts/s (Fiore et al. 1999).
The use of inter-calibration factors outside these ranges can affect the evaluation
of the actual spectral distribution, so these factors cannot be considered as completely
free parameters in the fitting procedure or to fix their values to adjust some results.
In particular, when a spectrum is intrinsically curved, like in the case of Mkn~501, 
an improper choice of the inter-calibration factors introduces a bias in the estimation 
of the spectral parameters. 
In the analysis of the 1997 observations of Mkn~501 Pian et al. (1998) 
took $f_{MP}$~=~0.75, just outside the lower end of the admissible range. 
This value was further reduced by 35\% by Krawczynski et al. (2002),
but this choice is simply not justified since it is not consistent with the MECS and PDS 
effective areas. 
In our evaluations of the spectral parameters, we verified 
that these two parameters were always inside the nominal ranges: we found
$f_{LM}$ in the range 0.75--0.85, while on some occasions $f_{MP}$ was found to be outside
the allowed interval and therefore we fixed it to an admissible value.
Applying a log-parabolic model, we obtained very satisfactory results for all 
the observations  with well acceptable $\chi^2$ values. The best fit values of these two
instrumental parameters and the resulting $\chi^2$ are given in Table 3.

\begin{table*}
\caption{ Best fit spectral parameters of the log-parabolic model for Mkn~501.}
\label{tab4}
\begin{tabular}{lccccccc}
\hline
Date & $a$ & $b$ & $K$ & $E_p$ (keV)  & $\nu_p F(\nu_p)^{(2)}$ & $F_{bol}^{(2)}$ & $F_{2-10~keV}^{(2)}$\\
\hline
1997-04-07 & 1.68 (0.01) & 0.17 (0.01) & 6.24 (0.08)~~10$^{-2}$ & 8.7 (1.3) & 1.41 (0.05)& 9.2 (0.4) & 2.15 (.01)\\
1997-04-11 & 1.64 (0.01) & 0.12 (0.01) & 6.09 (0.07)~~10$^{-2}$ & 31.6 (9.6) & 1.8 (0.1) & 14.1 (1.1) & 2.39 (.01)\\ 
1997-04-16 & 1.41 (0.01) & 0.147 (0.007)& 9.60 (0.1)~~10$^{-2}$ & 101.6 (23.7)& 6.0 (0.5) & 42.2 (3.5) & 5.24 (.01)\\
           &             &              &      &     &      &     &      \\
1998-04-28 & 1.65 (0.02) & 0.15 (0.02) & 4.74 (0.08)~~10$^{-2}$ & 14.7 (5.7) & 1.2 (0.1) & 8.4 (0.9) & 1.76 (.01) \\ 
1998-04-29 & 1.62 (0.02) & 0.17 (0.02) & 5.43 (0.08)~~10$^{-2}$ & 13.1 (4.3) & 1.4 (0.1) & 9.2 (0.9) & 2.08 (.01) \\
1998-05-01 & 1.71 (0.02) & 0.24 (0.02) & 4.77 (0.08)~~10$^{-2}$ & 4.0 (0.6) & 0.94 (0.03)& 5.1 (0.3) & 1.47 (.01) \\ 
1998-06-20 & 1.79 (0.02) & 0.21 (0.02) & 1.75 (0.04)~~10$^{-2}$ & 3.2 (0.5) & 0.32 (0.01)& 1.9 (0.1) & 0.495 (.005) \\ 
1998-06-29 & 1.69 (0.02) & 0.23 (0.02) & 3.23 (0.06)~~10$^{-2}$ & 4.7 (0.8) & 0.66 (0.03)& 3.7 (0.2) & 1.03 (.01) \\
1998-07-16 & 1.70 (0.02) & 0.33 (0.02) & 3.20 (0.07)~~10$^{-2}$ & 2.8 (0.3) & 0.60 (0.02)& 2.8 (0.1) & 0.90 (.01) \\
1998-07-25 & 1.76 (0.01) & 0.28 (0.01) & 3.37 (0.05)~~10$^{-2}$ & 2.7(0.1) & 0.61 (0.01)& 3.1 (0.1) & 0.93 (.01) \\
           &             &              &      &     &      &      &    \\
1999-06-10$^{(1)}$ & 2.15 (0.01) & 0.24 (0.01) & 1.86 (0.02)~~10$^{-2}$ & 0.49 (0.03) & 0.315 (0.004) & 1.73 (0.04) & 0.302 (.002) \\
\hline
\multicolumn{7}{c} { }
\end{tabular}
Errors are given at 1 sigma for one interesting parameter. \\
(1) ~~PDS data not included. \\
(2) ~~In units of 10$^{-10}$ erg cm$^{-2}$ s$^{-1}$. \\
\end{table*}

The best fit values of the spectral parameters for the log-parabolic model
are reported in Table 4. 
For each observation we report the values of $a$, $b$ and $K$, together with 
four derived parameters: the peak energy, the SED peak value, the bolometric flux 
of the considered spectral component and the 2--10 keV flux. 
Errors correspond to 1 standard deviation for one interesting parameter ($\Delta \chi^2$=1).

The three observations of April 1997 were performed during a very active phase of
Mkn~501. The peak energy $E_p$ changed from about 9 keV to 100 keV and
correspondingly the $F_{bol}$ changed by a factor of $\sim$4.5.
The curvature parameter $b$ was generally low, in the range 0.12--0.17.
The SEDs of these observations are shown in Fig.~2.
Despite the good $\chi^2$, the residuals showed some systematic deviations.
In particular, in the spectrum observed on April 11, there is a clear excess 
at energies above 30 keV and a similar behaviour can also be recognised in 
the data of April 7. 
To show in detail how this deviation is significant we plotted the best fit 
spectrum and residuals of April 11 in Fig.~3: the last seven points in the PDS 
range are all more than one standard deviation higher than expected.
Note also that if the best fit parameters are evaluated limiting the upper energy
to 30 keV, their values do not change indicating that their estimation is dominated 
by the low energy data because of their much better statistics.
Another important result, very clearly evident in the SEDs of Fig.~2, is that the
flux at energies below 0.5 keV remained practically unchanged despite the
large variability observed at higher energies. As discussed in Sect.~5, 
this fact can be relevant to look for a possible interpretation of the spectral 
evolution of the flare.

A marginal detection (1997, April 4-15) of Mkn~501 with EGRET at energies greater 
than $\sim 100$ MeV is reported by Kataoka et al.~(1999) who give the photon flux 
$F(>100 MeV) = 9\pm7 \times 10^{-8}$ ph cm$^{-2}$ s$^{-1}$. 
We verified that the extrapolation in this range of our log-parabolic fit for 
the April 16 observation, when the source was at the highest level in the X-ray band, 
was compatible with this flux. 
Indeed, using the best fit values of Table 4, we computed a flux $F(100 MeV) = 1.8 
\times 10^{-9}$ ph cm$^{-2}$ s$^{-1}$ MeV$^{-1}$ corresponding to a power law integral 
flux of $F(>100 MeV) \simeq 16 \times 10^{-8}$ ph cm$^{-2}$ s$^{-1}$. 
Note that this value, considering that the source during the EGRET pointing was likely 
weaker than the exceptional flare of April 16 and that its $\gamma$-ray spectrum steeper 
than our assumption, must be considered as an upper limit. 
In any case, the April 16 extrapolated value is compatible with the EGRET result 
within $1 \sigma$. 

Observations of Mkn~501 from March to October 1997 at energies higher than 0.25
TeV, performed with CAT, are reported by Djannati-Atai et al. (1999), who adopted
a log-parabolic law to fit the spectra. In particular, in April the source was 
observed in time windows very close, but not strictly simultaneous, with the 
{\it Beppo}SAX pointings. It showed a behaviour remarkably similar to that in 
the X rays with a very strong flare on April 16. The spectral curvature in the
TeV range was always stronger than in the X rays, in particular the value
of $b$ was found greater than 0.4. A similar result is reported by
Krennrich et al. (1999), who observed Mkn~501 with the Whipple telescope 
from February to June 1997.

\begin{figure*}[ht]
      \vspace{0.8cm}
      \hspace{2.0cm}
\epsfysize=9cm
\epsfbox{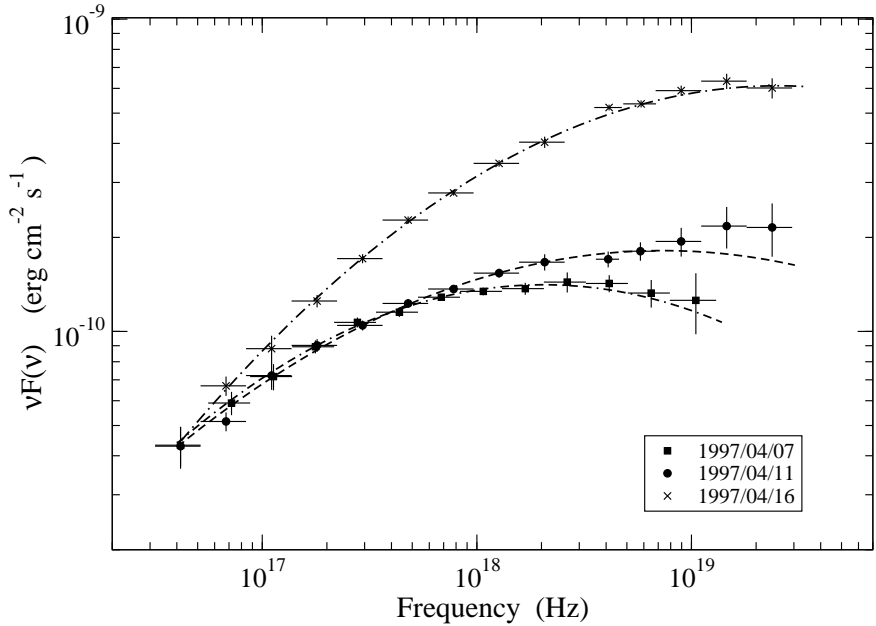}
\caption[]{
The X-ray Spectral Energy Distribution of Mkn 501 observed with {\it Beppo}SAX 
in April 1997. The interpolations are the best fits with a log-parabolic law.
}
%\label{S97M501}
\end{figure*}

The behaviour of Mkn 501 in the subsequent observations was different
from that of 1997. During the long campaign of spring-summer 1998 the source 
was observed several times: in the two observation of April 1998 it was rather
bright with a peak energy around 14 keV and a spectral curvature similar
to that measured the previous year. Starting from May 1998, $b$ increased to 0.2--0.3
and $E_p$ decreased to about 3--4 keV; such state remained rather stable up
to the last observation of July 25. The following year Mkn~501 was observed only once 
when $E_p$ dropped to about 0.5 keV, while the curvature was unchanged.
Some SEDs of these observations are plotted in Fig.~4. 
Note that in these observations, contrary to the SEDs of Fig.~2, there is 
a significant variability of the source flux below 0.5 keV.

\begin{figure}
      \vspace{-1.0cm}
\epsfysize=8cm
\epsfbox{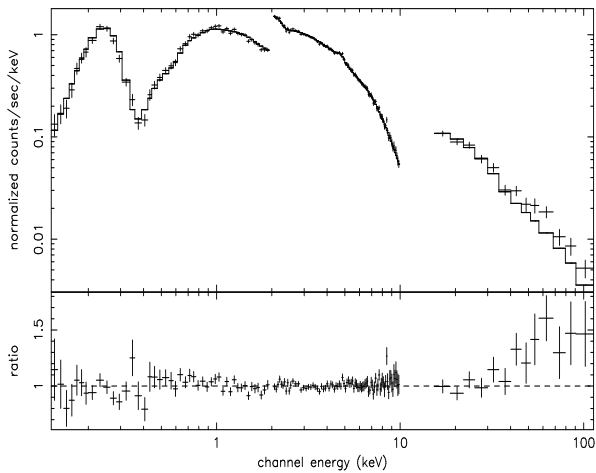}
\caption[]{
The best fit spectrum with a log-parabolic law and the residuals of the {\it Beppo}SAX 
observation of Mkn~501 on 11 April 1997.
The binning of the data is finer than in Fig.~2. Note the systematic deviation
due to a flux excess at energies greater than 30 keV.
}
%\label{S970411M501}
\end{figure}

\section{Optical observations}

Optical observations are useful to verify how and if the X-ray SEDs extend
to a much wider frequency interval and whether the log-parabolic model still fits the data 
even in this range.
A direct use of photometric data, especially when obtained with small aperture
telescopes, is complicated by the fact that the nuclear emission on Mkn~501 is out-shined
by the much brighter host elliptical galaxy.
To correct the multiband photometric data for the galaxian component it is necessary
to know how much it contributes to the total flux within the used aperture, although
other contributions in the optical range from the nuclear environment cannot be excluded.

\begin{figure*}[ht]
      \vspace{0.8cm}
      \hspace{2.0cm}
\epsfysize=9cm
\epsfbox{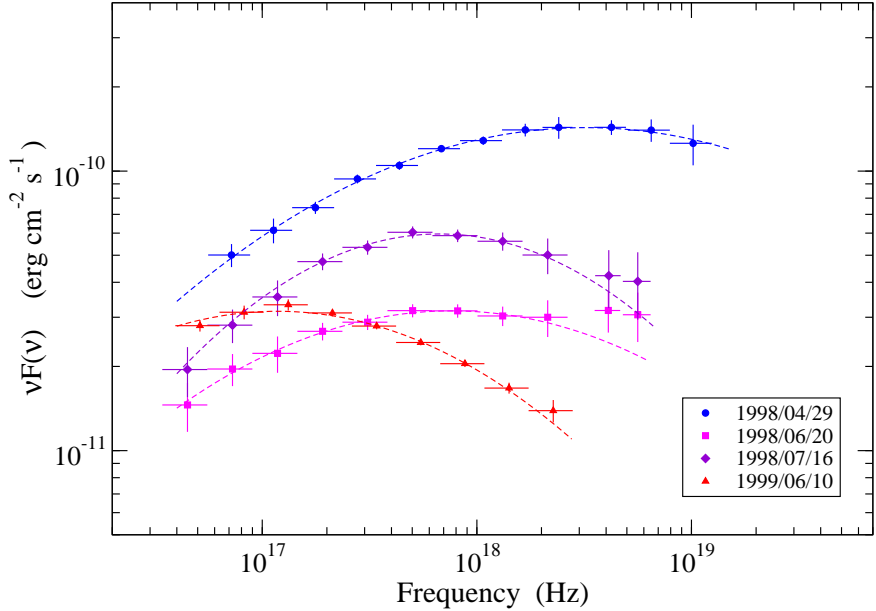}
\caption[]{
The X-ray Spectral Energy Distribution of Mkn 501 observed with {\it Beppo}SAX 
in 1998 and 1999. The interpolations are the best fits with a log-parabolic law.
}
%\label{S978M501}
\end{figure*}

We performed photometric observations of Mkn~501 in the $B$, $V$
(Johnson) and $R$, (Cousins) bandpasses on 26 May 1997, about one month
after the large outburst in the hard X rays, 17 and 22 June 1998, close to
{\it Beppo}SAX pointings, and on 23 June 2000. We used the 70 cm reflector telescope
of the University of Roma and IASF-CNR located at Monte Porzio and equipped with
a back-illuminated CCD camera (Site 501A). Standard stars were taken from
Gonzales-Peres et al. (2001).
The magnitudes of Mkn~501, computed with a photometric radius of 4$''$,
are given in Table 5. The variability is of the order of a tenth of a magnitude and
this quite small amplitude can be understood because the flux is dominated by the
galaxian contribution which is larger than the nuclear emission.
A photometric study of the host galaxy in the $R$ band has been
performed by Nilsson et al. (1999), from which one can estimate a magnitude
$R_{gal}$=13.90 for our photometric radius.
The magnitudes in the other two bands were estimated assuming typical colours 
for an elliptical galaxy (Fukugita et al. 1995) and resulted equal to $B_{gal}$=15.47
and $V_{gal}$=14.51. Of course, these estimates are affected by systematic
errors to be added to the measure uncertainties.
Fluxes were then computed using the zero magnitude values from Mead et al.~(1990) 
and the nuclear component was obtained after the
subtraction of the host galaxy contribution; the resulting values are
$F(R)$=5.01, $F(V)$=3.79 and $F(B)$=1.96 mJy.
We also assumed the extinction $A_V = 0.05$ in the direction of Mkn~501.

The resulting optical--X-ray SED for the observation performed on 20--21 June
1998 is shown in Fig.~5: we plotted the data before and after the subtraction 
of the host galaxy contribution to illustrate how much this affects the evaluation
of the SED. The resulting nuclear spectrum is very steep, although one cannot exclude
a further contribution from the circumnuclear environment or residual systematic effects.
In any case, as for Mkn~421, it is fairly evident that optical data cannot match the 
low frequency extrapolation of the X-ray spectrum, suggesting the presence of different
emission components.

The amplitude of optical variations of Mkn~501 is much smaller than that
measured in the X rays. As discussed above, the amount of nuclear flux largely
depends on the subtraction of the galaxian contribution within the considered 
photometric aperture and
therefore the actual luminosity changes of the nuclear emission are poorly known.
In any case, to verify if the X-ray emission in the faintest state is 
compatible with the optical data, we plotted in the Fig. 5 also the X-ray SED 
of Mkn~501 observed in June 1999 when the SED peak was shifted toward the
lower frequencies. We see that also in this case any reasonable extrapolation of 
the X-ray and optical data cannot match both the flux values and the spectral shape.

\begin{table}
\caption{ Optical photometric data of Mkn~501.}
\label{tab3}
\begin{tabular}{lccc}
\hline
Date & B & V & R \\
\hline
1997/05/26 &  &  & 13.26$\pm$0.02  \\
1998/06/17 &  &  & 13.40$\pm$0.02  \\
1998/06/22 & 14.83$\pm$0.03 & 13.89$\pm$0.02 & 13.32$\pm$0.02 \\
2000/06/23 & 14.62$\pm$0.02 & 13.99$\pm$0.02 & 13.48$\pm$0.02 \\
\hline
\multicolumn{4}{c} { }
\end{tabular}
\end{table}

\section{Discussion}

Spectral variability in blazars is a very complex issue since it 
involves rapidly evolving processes that depend on a large number of physical quantities. 
It is important, therefore, that spectral analysis makes use of models based on simple
analytical representations that can be directly related to some physical parameters.
In our analysis of the {\it Beppo}SAX X-ray observations of Mkn~501, covering more
than three decades in energy, we used a log-parabolic spectral model which
represents a step in this direction.
We have shown that the curved spectra of this source are generally well
represented by a law of this type, while a single power law with an exponential
cut-off or a broken power law generally do not give equally acceptable best fits.

\begin{figure}[ht]
      \hspace{-0.4cm} 
      \vspace{+0.6cm}
\epsfysize=7cm
\epsfbox{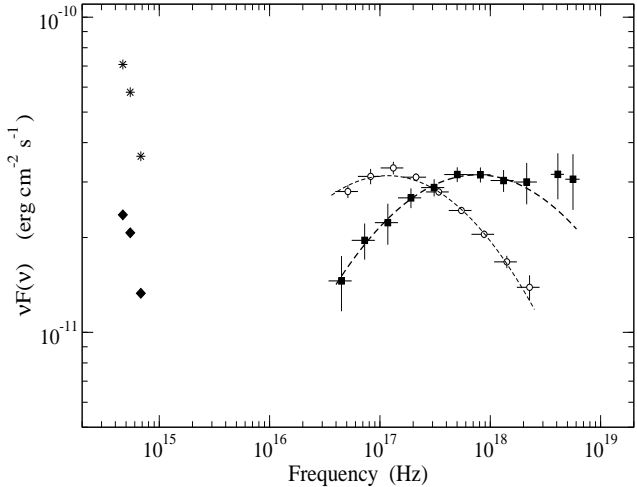}
\caption[]{
The optical to X-ray SED of Mkn~501 in June 1998. The X-ray data are those of 
the June 20--21 (filled squares) and the long-dashed line is the log-parabolic best
fit with the parameters given in Table 4. Optical data were obtained during a
nearly simultaneous observation (22 June 1998): stars correspond to the 
measured magnitudes, filled diamonds are the estimated nuclear SED after the 
subtraction of the host galaxy contribution as described in Sect.~4. For comparison
we plotted also the X-ray SED observed in June 1999 (open circles) when Mkn~501
was in a much fainter state.
}
%\label{S97M501pp}
\end{figure}

\begin{figure}[ht]
      \hspace{-0.2cm}
      \vspace{0.9cm}
\epsfysize=7cm
\epsfbox{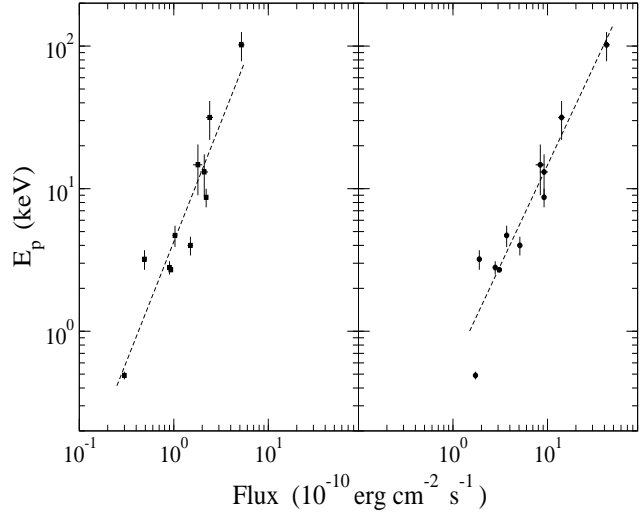}
\caption[]{
The correlations between the peak energy of the SEDs and the flux in the 2--10
keV (left panel) and the bolometric flux (right panel) estimated from Eq.(5).
Dotted lines are the power law best fits.
}
%\label{M501FEcor}
\end{figure}

The most common theoretical approach to produce curved spectra is to consider radiative losses
and the escape of high energy electrons from the emitting region: the literature
has been summarized in the recent paper by Krawczynski et al. (2002). The intrinsic
difficulty is that complex numerical calculations are necessary to solve the transfer
equation of the electrons and consequently it is hard to introduce a single parameter
to describe the spectral curvature and to find how this is related with the
other main physical parameters of the model.

As discussed in Paper I the log-parabolic spectral distribution can be related
to the particle acceleration mechanism. We have shown that such a law can be obtained
when the acceleration probability is a decreasing function of the particle
energy, approximable with a simple power relation, and the curvature parameter $b$
measures this effect.
The comparison of the results on Mkn~501 with those on Mkn~421 (Paper I) can be
useful to understand how this physical process works in these two sources.
As a first point we note that the $b$ values of Mkn~501 are generally smaller than
those of Mkn~421: for the latter source we found that $b$ was usually in the range
0.35--0.48 and decreased to $\sim$0.2 during the large outburst of May 2000.
The values given in Table 4 show that for Mkn~501 $b$ was typically 0.18--0.24 (only
in one case it increased to 0.33) and during the bright states it decreased
to $\sim$0.15. This result may indicate that the physical conditions in
Mkn~501 produce a lower decrease with energy of the particle acceleration probability 
than in Mkn~421.
For instance a larger volume or a higher magnetic field can make more efficient the
particle confinement. In this case one can also expect that energetic particle can
reach higher energies and this agrees with the higher peak energies observed in
Mkn~501. Note, however, that its integrated X-ray luminosity (assumed
proportional to $F_{bol}$) is comparable to Mkn~421: in particular in the brightest
states it was higher only by a factor of about 2.

As mentioned in Sect.~3 spectral curvature is also well evident in the TeV range,
where Djannati-Atai et al. (1999) and Krennrich et al. (1999) reported for Mkn~501 values
of $b$ in excess of 0.4. Furthermore, the spectrum of Mkn~421, observed by Whipple 
in May 1996, was characterised by a $b$ value of 0.28$\pm$0.09, significantly 
less curved than Mkn~501. X-ray observations of Mkn~421 in May 1996 performed by ASCA showed 
that the source had a high flaring activity and reached a 2--10 keV flux of $\sim$ 4 10$^{-10}$ 
erg cm$^{-2}$ s$^{-1}$ (Kataoka et al.~1999), comparable to that measured in spring 2000 
(see Paper I).
Unfortunately, measures of $b$ for that period are not available, but if the SED
was similar to that observed with {\it Beppo}SAX in May 2000 a value of $b$ of
$\sim$0.18--0.25 can be assumed.
Under this assumption we found that the X-ray and TeV spectral curvatures
of Mkn~421 can be rather similar and thus these emission can be related to the
same electron population.

\begin{figure*}[th]
      \vspace{0.cm}
      \hspace{2.0cm}
\epsfysize=8cm
\epsfbox{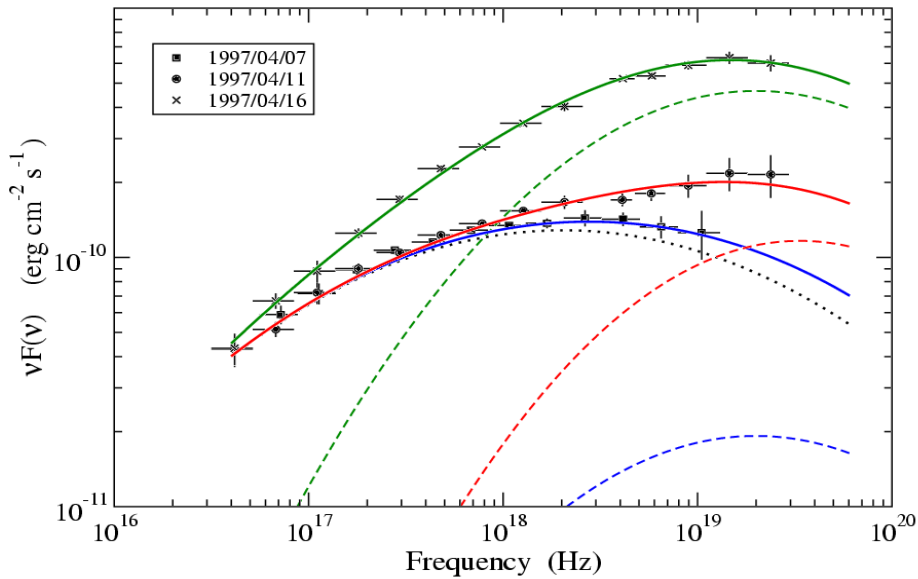}
\caption[]{
A two model component of the X-ray flare of Mkn 501 observed in April 1997.
The nearly constant component is the dotted line, while three states of
the variable component are represented by the dashed lines. Total SEDs are
the thick solid lines modelling the observed data.
}
%\label{S97M501pp}
\end{figure*}

The case of Mkn~501 looks different because in the X rays $b$ values as high as
that at TeV energies has never been measured. From Table 4 we see that the highest
curvature ($b\simeq0.3$) was found in 1998 when the source was rather faint, and in
1997 it was much brighter for a period lasting several months.
We do not know if these differences of the spectral curvature are genuine or
due to the superposition of multi-epoch TeV data. 
The more pronounced curvature of the TeV spectra could be produced by several processes like pair 
production by energetic photons against the infrared radiation in the intergalactic 
space or the inverse Compton scattering in the Klein-Nishina regime. In a further paper
(Massaro et al. - Paper III, in preparation) we will investigate, by means of a
numerical code, the curvature properties of the synchrotron and inverse 
Compton emission from an electron population with a log-parabolic energy distribution.

The log-parabolic model gives us also the possibility of investigating the correlation
between some spectral parameters like the peak energy and the flux.
Tavecchio et al. (2001) already searched for this correlation but their result was
not very stringent because they found many lower limits on $E_p$ and therefore were
only able to conclude that their analysis suggested a relation $E_p \propto F^n$
with $n \simeq 2$.
In Fig.~6 we plotted the same relation using the best fit values given in Table 4:
we used both the flux in the 2--10 keV band and the bolometric flux and found
very good correlations with $E_p$. The linear correlation coefficients of $Log~E_p$
with $Log~F_{2-10~\rm{keV}}$ and $Log~F_{bol}$ are 0.937 and 0.948, respectively,
confirming the existence of power relationships. The resulting exponents
of these two laws are $1.68\pm 0.21$ and $1.41\pm 0.16$, in particular the latter
significantly smaller than 2.
According to Tavecchio et al. (2001) the quadratic dependence of the peak energy
on the luminosity can be accounted by a simple scenario in which only the maximum
Lorentz factor of the emitting electron varies whereas the other physical quantities
remain almost constant.
Note that in Fig.~6, right panel, there is a point largely out of the correlation:
it corresponds to the long 1999 observation, in which the peak energy was very low
while the bolometric flux was practically equal to that of 20 June 1998, as also
shown by the SEDs in Fig.~4. This can be an indication that the physical conditions
in the nucleus of Mkn~501 at that epoch were really very different from those
of the preceding observations. Excluding this point, the correlation
coefficient improves to 0.963 and the value of $n$ changes to 1.25$\pm$0.12, even
smaller than 2.
We interpret this result as an indication that not only the maximum Lorentz factor
changed but other quantities should have varied as well.

The X-ray and TeV emission of Mkn~501 was explained by Tavecchio et al. (2001) with
a one-zone SSC model in which the X-ray SED changes are produced by a component
with a stable emission below a few keV and a variable high energy cutoff.
Krawczynski et al. (2002), however, found that a one-component model fails
to fit both the SED and the light curves at different energies from keV to TeV.
These authors obtained better fits considering a two-component model, one of which
in the spectral range 3--25 keV, comparable to our LE component (see below). However, they
did not show details of the X-ray SEDs and it is not possible to verify if the
adopted power law distribution can fit the spectral curvature below 10 keV.

We already pointed out in Paper I that the SED of Mkn~421 can be explained by means
of the simultaneous presence of different emission components. Likely, the same holds
for Mkn~501. Suggestive indications for the presence of more than one component in Mkn~501 are
derived from the spectral dynamics during the flare of April 1997.
As noted in Sect.~4 the flux at energies below $\sim$0.5 keV remained very stable
while that above $\sim$30 keV changed by about an order of magnitude.
In particular on April 11 we found evidence for a high energy excess with respect to
the log-parabolic best fit.
Furthermore, the optical spectrum is generally steep and does not match the low energy
extrapolation of the X-ray spectrum.
A better description of the observed spectra can be obtained adding a second emission
component: in the X-ray range one component (LE) dominates at
lower energies and its peak energy in the SED lies between 0.5 and $\sim$5 keV, while a 
second one (HE), responsible for the emission in the hard X rays, peaks
at energies higher than 10 keV and likely up to $\sim$150 keV and can show large
variations on short time scales. We assume that the physical mechanism for the particle 
acceleration and emission are basically the same and therefore the spectra of both components 
are given by log-parabolic laws. This does not mean that six parameters (instead of three) are 
{\it necessary} to fit the data. In fact, a best fit of the data using two log-parabolas 
is poorly determined and acceptable solutions are obtained in wide ranges of the parameter space. 
We then tried to reproduce the observed spectral shape constraining the model with some assumptions.
We assumed that the spectra of both components are described by log-parabolic laws
and choose the parameter of the LE component approximately equal to those of April 7
($a$=1.68, $b$=0.18, $K$=5.7$\times$10$^{-2}$, the corresponding SED is the dotted line in Fig.~7) 
allowing a moderate variability of $a$ to 1.60 and of $K$ to
6.9$\times$10$^{-2}$ on April 16.
The parameters of the HE component were established to match the sum of the components
and the data: on April 11 the values of $a$ and $b$ were 0.50 and 0.35 while on April 7
and 16 the were 0.85 and 0.35, respectively; the normalisation $K$ varied between
0.95$\times$10$^{-3}$ and 23$\times$10$^{-3}$.
The HE components and the total SEDs are plotted in Fig.~7:
the agreement is fully satisfactory and the excess above 30 keV
is well reproduced.
With these values of $a$ and $b$ the HE peak energy varied between 83 and 140 keV.
The actual spectral shape of the HE component at energies lower than a $\sim$1 keV, 
for instance a power law instead of a log-parabola, is not important because its
contribution to the total flux is about an order of magnitude smaller than
the LE component.
The X-ray spectral changes observed are therefore mainly produced by a variation of the
number of emitting particles of both components:
for the LE component the particle number increased of about 20\%, while for the HE
it increased by a factor of $\sim 25$.
Note that this two-component model does not require low $b$ values as those
found on April 11 and 16: the smooth SEDs with reduced curvature are obtained
by adding components with curvatures similar to those observed in 1998
when Mkn~501 was less bright. Furthermore, the similar $b$ values can be
considered as an indication that the particle acceleration does not occur in very
different physical conditions. As a final remark we note that these two 
component model does not affect the correlations of Fig.~6 because it changes only 
the two highest flux points. The HE component assumed by us does not follow 
this correlation being its peak energy practically constant despite the large flux 
change. However this component can be present only during the exceptional flare of April 1997 
and thus not represent the typical behaviour of the source.

\begin{acknowledgements}

The CNR Institutes  and the {\it Beppo}SAX Science Data Center
are financially supported by the Italian Space Agency (ASI) in the
framework of the {\it Beppo}SAX mission.
Part of this work was performed with the financial support Italian MIUR
(Ministero dell' Istruzione Universit\'a e Ricerca) under the grant
Cofin 2001/028773.

\end{acknowledgements}

\setcounter{table}{1}
\begin{table}
\caption{ {\it Beppo}SAX WFC observation log.}
\label{tab2}
\begin{tabular}{lrrcc}
\hline
Date & Exp. & Offset & Count Rate$^1$ & Flux$^2$ \\
%    &  (s)$~~~$      & (deg)    &  (cts/s)       & (erg\,cm$^{-2}$\,s$^{-1}$)  \\
    &  (s)$~~$      & (deg)    &  (cts/s)       &  (cgs) \\
\hline
1996/09/19 &  67,352  &  12.4  & 0.92$\pm$0.10 &  0.95$\pm$0.10 \\
1997/01/04 &  11,994  &  13.7  & 1.78$\pm$0.21 &  1.82$\pm$0.21 \\
1997/01/06 &  18,496  &   9.3  & 2.28$\pm$0.13 &  2.34$\pm$0.13 \\
1997/01/11 &  19,713  &   6.0  & 1.65$\pm$0.09 &  1.69$\pm$0.09 \\
	1997/01/27 &  16,370  & 17.0  & 1.78$\pm$0.23 & 1.82$\pm$0.24 \\
1997/01/29 &  34,958  &   8.6  & 2.10$\pm$0.09 &  2.15$\pm$0.09 \\
	1997/02/22 &   9,621  & 14.4  & 2.17$\pm$0.27 & 2.22$\pm$0.27 \\
	1997/03/02 &  18,414  &  1.8  & 2.35$\pm$0.09 & 2.41$\pm$0.09 \\
	1997/03/04 &  12,226  & 13.4  & 1.96$\pm$0.21 & 2.01$\pm$0.22 \\
 	1997/03/05 &  25,846  & 10.4  & 2.31$\pm$0.13 & 2.37$\pm$0.13 \\
1997/03/15 &  28,148  &  12.5  & 2.42$\pm$0.16 &  2.48$\pm$0.16 \\
1997/03/26 &  24,729  &  17.4  & 2.77$\pm$0.25 &  2.84$\pm$0.26 \\
1997/04/01 &  11,490  &  17.1  & 2.87$\pm$0.26 &  2.94$\pm$0.27 \\
1997/08/30 &  51,124  &  18.5  & 2.34$\pm$0.25 &  2.39$\pm$0.25 \\
1997/09/04 &  11,342  &   4.5  & 2.51$\pm$0.13 &  2.57$\pm$0.13 \\
1997/09/06 &  11,991  &   2.9  & 2.37$\pm$0.11 &  2.42$\pm$0.11 \\
	1997/10/01 &  17,069  &  2.9  & 2.52$\pm$0.10 & 2.58$\pm$0.10 \\
1997/10/02 &  11,121  &   4.5  & 3.16$\pm$0.13 &  3.24$\pm$0.14 \\
1998/02/02 &   8,256  &  12.4  & 1.50$\pm$0.20 &  1.53$\pm$0.20 \\
1998/02/24 &  28,014  &   3.9  & 1.13$\pm$0.08 &  1.16$\pm$0.08 \\
1998/02/28 &  12,388  &   3.9  & 1.71$\pm$0.10 &  1.75$\pm$0.10 \\
1998/03/08 &  35,681  &  13.1  & 1.55$\pm$0.17 &  1.59$\pm$0.17 \\
1998/07/02 &  25,965  &   9.0  & 1.02$\pm$0.13 &  1.05$\pm$0.14 \\
1998/07/29 &  96,891  &  13.3  & 1.53$\pm$0.10 &  1.56$\pm$0.10 \\
1998/08/06 &  34,447  &   3.6  & 1.44$\pm$0.07 &  1.48$\pm$0.07 \\
1998/08/11 &  27,383  &   4.2  & 1.19$\pm$0.08 &  1.22$\pm$0.08 \\
1998/08/24 &  19,363  &   6.2  & 1.47$\pm$0.08 &  1.51$\pm$0.08 \\
1998/09/03 &  11,044  &   8.5  & 0.95$\pm$0.14 &  0.97$\pm$0.15 \\
1998/09/04 &  18,926  &   8.5  & 1.50$\pm$0.13 &  1.53$\pm$0.13 \\
1998/09/16 &  24,681  &   3.6  & 1.22$\pm$0.06 &  1.25$\pm$0.07 \\
1998/09/17 &  14,362  &   6.2  & 1.61$\pm$0.14 &  1.65$\pm$0.14 \\
1998/10/07 &  18,958  &  12.4  & 1.22$\pm$0.16 &  1.25$\pm$0.17 \\
1998/10/27 &  16,099  &   9.2  & 1.18$\pm$0.11 &  1.20$\pm$0.11 \\
1999/02/07 &  55,780  &  19.9  & 1.92$\pm$0.25 &  1.96$\pm$0.26 \\
1999/02/27 &  22,592  &   7.5  & 2.11$\pm$0.10 &  2.16$\pm$0.10 \\
1999/03/14 &  61,290  &  12.2  & 1.23$\pm$0.10 &  1.26$\pm$0.10 \\
1999/03/29 &  74,910  &   6.6  & 1.05$\pm$0.07 &  1.08$\pm$0.07 \\
1999/03/31 &  50,662  &  10.4  & 1.02$\pm$0.07 &  1.05$\pm$0.07 \\
	1999/04/05 &  30,384  & 16.6  & 1.41$\pm$0.16 & 1.45$\pm$0.16 \\
1999/05/01 &  53,604  &  14.3  & 0.81$\pm$0.12 &  0.83$\pm$0.12 \\
1999/08/08 &  62,916  &   3.6  & 0.41$\pm$0.05 &  0.42$\pm$0.05 \\
1999/08/13 &  18,946  &   2.9  & 0.81$\pm$0.10 &  0.83$\pm$0.10 \\
	1999/08/28 &  61,318  & 13.7  & 1.08$\pm$0.17 & 1.10$\pm$0.18 \\
	1999/09/11 &  23,178  &  6.1  & 0.95$\pm$0.10 & 0.97$\pm$0.10 \\
1999/09/24 &  15,640  &   8.9  & 1.09$\pm$0.13 &  1.12$\pm$0.13 \\
2000/02/15 &  32,282  &   3.9  & 0.46$\pm$0.07 &  0.47$\pm$0.07 \\
2000/03/05 &  24,423  &  13.3  & 0.92$\pm$0.13 &  0.95$\pm$0.13 \\
	2000/03/08 &  23,440  & 13.3  & 0.90$\pm$0.14 & 0.92$\pm$0.14 \\
2000/03/10 &  39,579  &  13.9  & 1.15$\pm$0.11 &  1.18$\pm$0.12 \\
2000/03/12 &  43,738  &   4.0  & 0.74$\pm$0.07 &  0.76$\pm$0.08 \\
2000/03/13 &  22,417  &   4.0  & 0.76$\pm$0.10 &  0.77$\pm$0.10 \\
	2000/03/19 &  22,013  &  2.6  & 0.69$\pm$0.09 & 0.70$\pm$0.09 \\
	2000/03/20 &  31,737  & 10.4  & 0.60$\pm$0.11 & 0.62$\pm$0.12 \\
2000/08/01 &  36,511  &   7.4  & 0.81$\pm$0.10 &  0.83$\pm$0.10 \\
	2000/08/17 &  85,411  &  4.5  & 0.42$\pm$0.05 & 0.43$\pm$0.05 \\
2000/08/20 &  33,638  &   6.9  & 0.66$\pm$0.09 &  0.67$\pm$0.10 \\
	2000/08/21 &  28,547  & 11.1  & 0.62$\pm$0.10 & 0.63$\pm$0.11 \\
2000/09/05 &  46,527  &   3.6  & 0.63$\pm$0.06 &  0.65$\pm$0.06 \\
2000/09/10 &  37,925  &   2.9  & 0.59$\pm$0.07 &  0.60$\pm$0.07 \\
\hline
\multicolumn{5}{c} { }
\end{tabular}
\end{table}
\setcounter{table}{1}
\begin{table}
%\caption{ {\it Beppo}SAX WFC observation log (cont.).}
\vspace{0.8 cm}
\label{tab2}
\begin{tabular}{lrrcc}
\hline
Date & Exp. & Offset & Count Rate$^1$ & Flux$^2$ \\
    &  (s)$~~$      & (deg)    &  (cts/s)       &  (cgs) \\
\hline
2000/09/23 &  39,293  &   7.6  & 0.67$\pm$0.09 &  0.69$\pm$0.09 \\
2000/10/03 &  37,820  &   6.9  & 0.91$\pm$0.08 &  0.93$\pm$0.08 \\
	2001/02/11 &  27,294  & 11.1  & 0.76$\pm$0.13 & 0.77$\pm$0.13 \\
2001/02/16 &  30,427  &  12.4  & 0.76$\pm$0.11 &  0.77$\pm$0.11 \\
	2001/02/25 &  33,444  &  2.6  & 0.64$\pm$0.07 & 0.66$\pm$0.07 \\
2001/08/09 &  37,884  &   4.5  & 0.55$\pm$0.07 &  0.56$\pm$0.07 \\
2001/08/14 &  81,153  &   8.6  & 0.63$\pm$0.07 &  0.65$\pm$0.08 \\
2001/08/17 & 120,472  &  10.4  & 0.82$\pm$0.08 &  0.85$\pm$0.08 \\
2001/08/22 &  56,668  &  10.0  & 0.63$\pm$0.09 &  0.65$\pm$0.09 \\
2001/09/11 &  37,706  &   0.3  & 0.71$\pm$0.07 &  0.73$\pm$0.07 \\
2001/09/21 &  58,224  &   0.3  & 0.55$\pm$0.05 &  0.56$\pm$0.05 \\
2001/10/03 &  23,808  &   0.3  & 0.64$\pm$0.07 &  0.66$\pm$0.07 \\
\hline
\multicolumn{5}{c} { }
\end{tabular}

$^1$ 2--8 keV energy band, i.e. the band of maximum sensitivity for the two instruments.

$^2$ 2--10 keV energy band, in units of $10^{-10}$ erg\,cm$^{-2}$\,s$^{-1}$.
\end{table}

\end{document}